\documentstyle[prabib,multicol,aps,epsfig]{revtex}

\newcommand{\be}{\begin{equation}}
\newcommand{\ee}{\end{equation}}
\newcommand{\bey}{\begin{eqnarray}}
\newcommand{\eey}{\end{eqnarray}}
\newcommand{\ra}{\rangle}
\newcommand{\la}{\langle}

\begin{document}
\draft

 \title {
 Crossover of quantum Loschmidt echo from golden rule decay to perturbation-independent decay
 }

\author{Wen-ge Wang and Baowen Li}

\address{
 Department of Physics, National University of Singapore, 117542
Singapore}

\date{\today}

\maketitle

\begin{abstract}

 We study the crossover of the quantum  Loschmidt echo (or fidelity)
 from the golden rule regime to the perturbation-independent exponential decay
 regime by using the kicked top model. It is shown that the deviation
 of the perturbation-independent decay of the averaged fidelity from the Lyapunov decay
 results from quantum fluctuations
 in individual fidelity, which are caused by the coherence in
 the initial coherent states.
 With an averaging procedure suppressing the quantum fluctuations effectively,
 the perturbation-independent decay is found to be close to
 the Lyapunov decay.
 We also show that the Fourier transform of the fidelity is determined directly by
 the initial state and the eigenstates of the Floquet operators
 of the two classically chaotic systems concerned.
 The absolute value part and the phase part of the Fourier transform of the fidelity
 are found to be divided into several correlated parts,
 which is a manifestation of the coherence of the initial coherent state.
 In the whole crossover region, some important properties of the fidelity,
 such as the exponent of its exponential decay
 and the short initial time within which the
 fidelity almost does not change,
 are found to be closely related to
 properties of the central part of its Fourier transform.

\end{abstract}


\begin{multicols}{2}

 \section{Introduction}

 The quantum Loschmidt echo, $M(t)$, measures the overlap of the evolution of the
 same initial state under two slightly
 different Hamiltonians in the classical limit,
 \be M(t) = |\la \Phi_0|{\rm exp }(iHt) {\rm exp}(-iH_0t) |\Phi_0 \ra |^2. \label{Mt}
 \ee
 Here $H_0$ is the Hamiltonian of a classically chaotic system and
 \be H=H_0 + \kappa V , \label{H} \ee
 with $\kappa$ being a small quantity.

 This quantity characterizes the stability of quantum motion under
 a small change of the Hamiltonian, e.g. by the interaction with
 the environment, it is thus called ``fidelity'' and of great
 interest in the fast developing field of quantum
 information\cite{Nielsen-C,Prosen01}.

 Moreover, a relationship between this quantity and the Lyapunov
 exponent that characterizes the classical chaos has been
 established analytically by using the semiclassical theory by
 Jalabert and Pastawski \cite{JP01}, which has been confirmed
 numerically in several models
 \cite{JSB01,CPW02,CLMPV02,WVPC02,WC02,BC02}. Therefore, this
 quantity has attracted a great attention from the community of
 quantum chaos \cite{Peres84,CT02,Prosen02,PZ02},

 It has been shown that several regimes exist.
 In the golden rule regime of the perturbation parameter, above a perturbative border,
 the fidelity has been found to have a simple decaying behavior,
 $M(t) \propto {\rm exp}(-\Gamma_L t) $ \cite{JSB01,CLMPV02,JAB02},
 where $\Gamma_L $ is the half width of
 the local density of states (LDOS) $\rho_L$,
 which has a Lorentzian form (Breit-Wigner form),
 \be \rho_L(E) = \frac{\Gamma _L/ 2\pi }{E^2 + \Gamma_L^2/4}, \label{BW} \ee
 with $\Gamma_L = 2\pi \rho U^2$.
 Here $\rho$ is the density of states and $U$ is the typical
 transition matrix element of $\kappa V$ between eigenstates of $H_0$.
 As the perturbation parameter $\kappa $ is increased,
 the exponential decay of $M(t) $ deviates from $ {\rm exp}(-\Gamma_L t) $,
 when $\Gamma_L $ becomes comparable to the band width of $H_0$
 and $\rho_L$ deviates from the Lorentzian form notably.
 A decay with the Lyapunov exponent $\lambda $ of the underlying classical chaotic dynamics,
 $ M(t) \propto {\rm exp}(-\lambda t) $, can appear in this regime of
 the perturbation parameter,
 when the (effective) Planck constant is small enough
 and the initial state is chosen suitably, e.g., a narrow wavepacket  \cite{JP01}.
 An interesting feature of this decay of $M(t)$
 is that it is perturbation-independent, in the sense that
 it is irrelevant to the strength of the perturbation $\kappa V$
 and is determined by the classical behavior of the system $H_0$
 (and also that of $H$ due to the smallness of the perturbation in the classical limit).
 Based on the semiclassical theory and the random matrix theory analysis,
 a transition of $M(t)$ from the $\Gamma_L$-decay to the $\lambda $-decay
 has been conjectured to occur
 at $\Gamma_L = \lambda $  \cite{JSB01}.
 Numerical results in several models support the conjecture \cite{JSB01,WVPC02,BC02}.

 The validity of results of the semiclassical theory
 depends on the value of the (effective) Planck constant.
 When the (effective) Planck constant is not small enough,
 the validity should be checked by direct quantum mechanical
 calculations. In fact,
 the behavior of the fidelity in this case is still not quite clear.
 For example, whether there is a sharp transition from
 the $\Gamma_L$-decay to the $\lambda $-decay,
 or the transition occur in a finite regime of the perturbation parameter.
 It is even unclear whether $M(t)$ could decay with the Lyapunov exponent.
 Indeed, a perturbation-independent but slower than ${\rm exp}(-\lambda t)$ decay
 has been observed in the kicked top model \cite{JSB01},
 the mechanism of which is still not clear.
 Meanwhile, the random matrix theory seems unsuitable for such problems,
 since perturbation-independent decay of $M(t)$ is usually
 initial-state-dependent,
 the feature of which is hard to be captured by the random matrix theory treatment.

 The quantity used in this paper in analyzing properties of
 the fidelity $M(t)$ is its Fourier transform, denoted by $F_M(E)$ in what
 follows. As will be shown in Sect.~\ref{F_M} that $F_M(E)$
 is determined directly by properties of the initial state and
 of the (quasi)energy eigenstates of the two classically chaotic systems concerned.
 The role played by $F_M(E)$ in understanding the behavior of $M(t)$
 is similar to that by the LDOS for the survival probability
 $P(t) = |\la \alpha |{\rm exp }(-iHt) |\alpha \ra |^2 $.

 In this paper, properties of $F_M(E)$ and their relations to those of $M(t)$ will be
 investigated numerically in the kicked top model \cite{Haake}.
 The fidelity in this model has been studied in Refs.~\cite{JSB01,JAB02,PZ02} and
 here we will concentrate on issues unaddressed in the previous work.
 Specifically, in addition to properties of $M(t)$ related to the function $F_M(E)$,
 we focus ourself on the following problems:

 (i) The mechanism of the perturbation-independent decay slower
 than ${\rm exp}(-\lambda t)$.

 (ii) Due to the small difference between $H_0$ and $H$ in the classical limit,
 the classical counterpart of the fidelity $M(t)$ should change quite slowly
 when $t$ is small enough (see Ref.~\cite{BC02} for numerical results in another model).
 In the quantum mechanical case,
 a similar phenomenon should be of practical interest in quantum computing.
 However, it was not clear whether or not the phenomenon exists.

 The paper is organized as the follows.
 The function $F_M$ is introduced in Sect.~\ref{F_M}.
 In Sect.\ref{numerical}, we shall present our numerical investigation
 of the problems mentioned above, in particular,
 for properties of the amplitude and the phase of the function
 $F_M(E)$ with initial coherent states.
 The perturbation-independent decay slower than ${\rm exp}(-\lambda t)$
 will be shown to be due to large quantum fluctuations in individual fidelity,
 which are caused by the coherence possessed the initial coherent states,
 when the (effective) Planck constant is not small enough.
 It will be shown that the fidelity $M(t)$ does not decay obviously
 within a short initial time,
 the length of which is determined by some properties of
 the function $F_M(E)$.
 The relationship between the decaying rate of $M(t)$ after the short initial time
 and some properties of the function $F_M(E)$ will also be shown numerically.
 Conclusions and discussions will be given in Sect.~\ref{Conclusion}.

\section {Function $F_M(E)$ as the Fourier transform of $M(t)$ }
\label{F_M}

 In this paper, the eigenstates of the two Hamiltonians $H_0$ and $H$
 are denoted by $|\alpha \ra $ and $|\beta \ra $,
 respectively, with eigenenergies $E_{\alpha}$ and $E_{\beta}$,
 \be H_0  | \alpha \ra = E_{\alpha} | \alpha \ra , \hspace{0.5cm}
 H| \beta \ra = E_{\beta} | \beta \ra . \label{Eab} \ee
 The expanding coefficient of $|\beta \ra $ in $|\alpha \ra $ is indicated by $C_{\beta \alpha}$,
 $C_{\beta \alpha} = \la \alpha | \beta \ra $.
 The LDOS $\rho_L(E)$ of an eigenstate $|\alpha \ra $ of $H_0$
 is determined directly by the eigensolutions
 the two Hamiltonians concerned,
 \be \rho_L(E) = \sum_{\beta} |C_{\beta \alpha}|^2 \delta
 \left ( E-(E_{\beta} - E_{\alpha}) \right ). \label{LDOS} \ee
 As is known, when the initial state $|\Phi_0 \ra $ is an eigenstate of $H_0$,
 the fidelity amplitude  $m(t)$,
 \be m(t) = \la \Phi_0|{\rm exp}(iHt) {\rm exp}(-iH_0t) |\Phi_0 \ra , \label{mat} \ee
 is just the Fourier transform  of the LDOS $\rho_L(E)$,
 and the form of the LDOS is important in the study of $M(t)$,
 which is just the survival probability $P(t)$ in this case.
 However, as shown in Ref.~\cite{WC02}, the LDOS can not explain the
 perturbation-independent decay of $M(t)$.
 In fact, the perturbation-independent decay of $M(t)$ appears only for some special
 set of initial states, that is, it is initial-state-dependent.
 Although a  rigorous condition is still lacking,  a sufficient condition for such
 initial states is expressed in a rough way that an initial state should be a
 narrow wavepacket\cite{JP01}.
 When the decay of $M(t)$ is initial-state-dependent,
 one should use a quantity that is more general than the LDOS
 in analyzing properties of $M(t)$.

 The function $F_M$,  the Fourier transform of the fidelity $M(t)$,
 is a suitable candidate, which is introduced as the follows.
 Let us first express the fidelity amplitude $m(t)$ in the form
 \be m(t) = \int f_m(\epsilon) e^{-i\epsilon t} d\epsilon, \label{fm3} \ee
 where
 \bey f_m(\epsilon) =  \sum_{\alpha ,\beta } A_{\alpha \beta } \delta
 \left [ \epsilon- (E_{\alpha}-E_{\beta}) \right ], \label{fma}
 \\ A_{\alpha \beta} = \la \Phi_0 | \beta \ra  \la \alpha |  \Phi_0\ra  C_{\beta \alpha }^*.
  \label{Aab} \eey
 Note that the quantity $A_{\alpha \beta}$ defined in Eq.~(\ref{Aab})
 is invariant under the transformation
 \be |\alpha \ra \to e^{i \varphi_{\alpha}} |\alpha \ra ,
 \hspace{0.2cm } |\beta \ra \to e^{i \varphi_{\beta}} |\beta \ra ,
 \hspace{0.2cm}  |\Phi_0 \ra \to e^{i \varphi_{0}} |\Phi_0 \ra , \label{trans} \ee
 where $  \varphi_{\alpha} ,\varphi_{\beta} $, and $  \varphi_{0}$ are arbitrary phases.
 That is, it is irrelevant to the relative phases among the states
 $|\alpha \ra $, $|\beta \ra $, and $|\Phi_0 \ra $.
 The fidelity $M(t)$ can be expressed as
 \be M(t) = |m(t)|^2 = \int F_M(E) e^{-iEt} d E, \label{Mt-FM} \ee
 where
 \be F_M(E) = \int f_m(\epsilon) f_m^*(\epsilon-E) d\epsilon. \label{FM} \ee
 The quantity $F_M(E)$ is in fact the correlation function of $f_m(\epsilon)$.
 Note that $F_M(E)$ can be calculated directly from properties of the
 states $|\alpha \ra $, $|\beta \ra $, and
 the initial state $|\Phi_0 \ra $.
 In the following section, we will investigate numerically the
 relationship between properties of $M(t)$ and those of $F_M(E)$.
 To this end, it is convenient to  write $F_M(E)$ as
 \be F_M(E) = \rho_F(E) e^{i\theta_F(E)}.  \label{rtF} \ee

 Some arguments can be given to the reason why the random matrix theory
 can be used to predict the decaying behavior of the fidelity $M(t)$
 in the golden rule regime,
 but not in the regime where the
 perturbation-independent decay of $M(t)$ appears.
 The initial states, whose fidelity is found to have perturbation-independent
 exponential decay controlled by the Lyapunov exponent,
 such as narrow wave packets in the configuration space \cite{JP01}, coherent states \cite{JSB01},
 momentum eigenstates \cite{BC02}, etc.,
 usually have wide spreading in the chaotic states $|\alpha \ra $ ($|\beta \ra $).
 Such initial states, which are regular in the classical limit, possess certain kind of coherence.
 It must be pointed out that the semiclassical theory does not predict the perturbation-independent decay of the fidelity
 for an arbitrary initial state \cite{JAB02}.
 As shown in the previous paragraph,
 the function $F_M(E)$, equivalently, $M(t)$, is basically determined
 by the quantity $A_{\alpha \beta }$ in Eq.~(\ref{Aab}).
 In the golden rule regime, the LDOS $\rho_L(E)$ is narrow, i.e.,
 $C_{\beta  \alpha } $ has many small components;
 as a result, for the quantity $A_{\alpha \beta }$, components of the initial
 states in the chaotic states,
 namely, $\la \Phi_0 | \beta \ra  \la \alpha |  \Phi_0\ra $,
 are not equally effective in the whole (quasi)energy region of $H_0$ and $H$.
 This should suppress the coherence of the initial states,
 which makes it possible to treat the components
 of the initial state taking part in the evolution of $M(t)$ effectively as random
 numbers, like
 in the random matrix theory treatment in Refs.~\cite{JSB01,CLMPV02,JAB02}.
 On the other hand, when the perturbation is so strong that
 the width of the LDOS is comparable to the (quasi)energy band width of $H_0$,
 the coherence possessed by the initial states should play a non-negligible
 role in the evolution of the fidelity.
 The random matrix theory can not be applied in this case,
 since it cannot describe the coherence in the initial states.

\section{ Numerical Study of Fidelity and Its Fourier Transform }
\label{numerical}

 \subsection{ The model }
 \label{model}

 The unperturbed Hamiltonian $H_0$ of the kicked top model used in this paper is
 \be H_0 = \frac{\pi }{2 \tau } S_y + \frac{K}{2S} S_z^2
    \sum_n \delta (t-n \tau ), \label{H0} \ee
 where $S$ is the total angular momentum, $\tau $ is the period,
 and $K$ is a parameter adjusting the strength of the kicks.
 Without loss of generality, the period $\tau $ is set to be unit, $\tau =1$.
 The model describes a vector spin that undergoes a free precession
 around the $y$-axis and is periodically perturbed by kicks
 around the $z$-axis.
 The time evolution of an initial state $|\Phi_0 \ra $ at $t=0$ is governed by the
 Floquet operator
 \be F_0 = {\rm exp} [ -i \frac{K}{2S} S_z^2] {\rm exp}
     [ -i \frac{\pi }2 S_y ], \label{F0} \ee
 where $\hbar $ has been set to be unit.
 The classical limit of the system, which is obtained by letting $S \to \infty $
 with $1/S$ serving as the effective Planck constant,
 is fully chaotic for $K\gtrsim 9$.

 The perturbation $\kappa V$ is chosen in the same way as in Ref.~\cite{JSB01},
 i.e., a slightly delayed periodic rotation of constant angle around the $x$-axis,
 \be V= \frac{\pi }{2\tau } S_x \sum_n \delta (t-n \tau -\epsilon ).
 \label{V} \ee
 The Floquet operator of the system $H$ is
 \be F = {\rm exp} [ -i \kappa \frac{\pi }2 S_x ] F_0, \label{F} \ee
 which gives the time evolution of an initial state,
 \be |\Phi(t) \ra = F^t |\Phi_0 \ra , \label{Phit} \ee
 where $t=0,1,2,\ldots $.
 For this model, discussions in Sect.~\ref{F_M} are still valid,
 with the eigenstates of $H_0$ and $H$
 changed to those of the Floquet operators $F_0$ and $F$, respectively,
 with the corresponding eigenenergies changed to quasi-energies.

 In numerical calculations, for the sake of convenience,
 the perturbation parameter $\kappa $ is written as
  \be \kappa = j \times 10^{-3}. \label{kappaj} \ee
 Unless addressed explicitly, the value of $S$ is 500
 for the numerical calculations discussed in this paper.
 Since the time $t$ in the evolution equation (\ref{Phit})
 takes integer values only,
 according to Eq.~(\ref{Mt-FM}),
 the domain of the variable $E$ in the function $F_M(E)$,
 which is $[-4\pi , 4\pi ]$ in the general case,
 can be reduced to $[-\pi , \pi ]$,
 with the definition of $F_M(E)$ changed accordingly,
 namely,
 \be F_M^{\rm fold}(E) = \sum_n F_M(E+2n\pi ),  \label{FMchange} \ee
 where $n$ takes the possible values in $(0, \pm 1, \pm 2 )$.
 The function $F_M^{\rm fold }(E)$
 is more closely related to properties of $M(t)$ than
 the basic one in Eq.~(\ref{FM}).
 It is this function that will be used in what follows in the numerical investigation of
 the kicked top model and,
 for brevity, it will be denoted  by $F_M(E)$.

 Initial states studied in this paper are in most cases coherent states of the SU(2) group
 \cite{Perelomov,Hecht},
 \bey |z\ra = A e^{z^* S_+}|M=-S \ra \label{CS}
 \\ z=-tan \frac{\theta }2 e^{i\phi }, \label{z} \eey
 where $|M\ra $ is the eigenvector of $S_z$,
 with the eigenvalue $M$ ($M=-S,-S+1, \ldots , S$),
 and $A$ is the normalization coefficient.
 The coherence in a coherent state $|z \ra $
 can be easily seen when it is expanded in the states $|M \ra $.
 However, in the study of the fidelity,
 $|z \ra $ is required to be expanded in the eigenstates of
 the Floquet operators $F_0$ and $F$.
 Although the expanding coefficients in such chaotic states must contain
 information on the coherence possessed by the coherent state,
 manifestation of the information is not so easy.
 In Sect.~\ref{numericalF_M}, we will show numerically that
 the function $F_M(E)$ introduced in Sect.~\ref{F_M}
 can supply some of the information.

\subsection{ Perturbation-independent decay of fidelity and quantum fluctuations }
 \label{PIdecay}

\begin{figure} 
 \epsfxsize=8cm
 \epsfbox{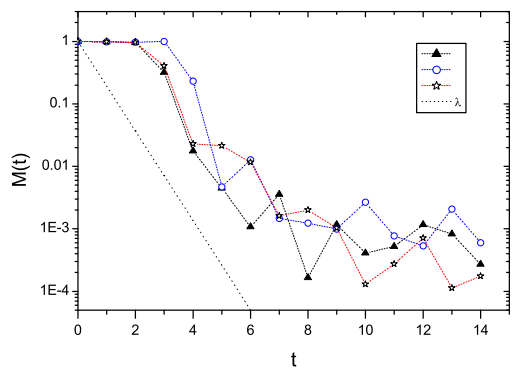}
 \vspace{-0.4cm}
 \narrowtext \caption{ \label{fig-singleMt} The fidelity $M(t)$ of three initial coherent states
 chosen arbitrarily,
 when $K=13.1$ and $j=6$.
 The dotted straight line shows the exponential decay with the corresponding Lyapunov exponent
 $\lambda =1.65$. }
 \end{figure}

 Basic properties of the fidelity $M(t)$ in the kicked top model
 have been studied in Ref.~\cite{JSB01}.
 Here, as mentioned in the introduction, we are interested in a perturbation-independent decay
 of the averaged $M(t)$ with initial coherent states,
 which decays more slowly than the exponential decay predicted by the semiclassical theory.
 In order to understand this phenomenon, we study the individual $M(t)$.
 In the regime of the perturbation parameter
 where the perturbation-independent decay appears,
 individual $M(t)$ is found to  have large fluctuations.
 See Fig.~\ref{fig-singleMt} for some examples of  $K=13.1$ and $j=6$,
 where $M(t)$ are plotted in the logarithm scale.
 On the other hand, in the golden rule regime, e.g., $K=13.1$ and $j=1$,
 no obvious fluctuation in individual $M(t)$ is observed.
 The large fluctuations are in fact
 quantum effects caused by the coherence in the coherent states.
 According to the arguments given in Sect.~\ref{F_M},
 they should appear when the expanding coefficients of the states $|\beta \ra $
 in $|\alpha \ra $ spread in the whole quasi-energy band of $F_0$ effectively,
 that is, when the half width of the LDOS becomes comparable to the bandwidth of
 the quasi-energy,
 which is found numerically for $j \ge 3$.
 In the classical limit $S \to \infty $,
 the fluctuations should disappear.

 As shown in Fig.~\ref{fig-singleMt},
 each $M(t)$ has a `shoulder' in a short initial time,
 denoted by $t_s$ in what follows, within which it changes slowly.
 The exponential-type decay of $M(t)$ appears only after $t_s$.
 In order to study the average decaying behavior of $M(t)$,
 their `shoulders' should be subtracted.
 To this end, we shift the time variable $t$ to
 $t_d=t-t_s$ and take average over $M(t_d)$.
 Here we meet a problem, since individual $M(t)$ has large fluctuations.
 The averaged $M(t_d)$ may be mainly determined by a small fraction of the $M(t_d)$ taken
 for averaging, if they have extraordinarily large values.
 Therefore, in addition to the standard averaged $M(t_d)$,
 denoted by $M_1(t_d)$,
 \be  {M_1}(t_d) =   \frac 1N \sum_{i=1}^N M(t_d,\Phi_i) , \label{aMt1} \ee
 we also calculate another averaged $M(t)$, denoted by $M_2(t_d)$,
 by taking the logarithm of $M(t_d)$ before performing the summation,
 \be \ln {M_2}(t_d) = \frac 1N \sum_{i=1}^N \ln M(t_d, \Phi_i) . \label{aMt2} \ee
 Here $M(t,\Phi_i)$ is the fidelity of an initial coherent state $|\Phi_i \ra $
 defined in Eq.~(\ref{Mt}), with the dependence on
  $|\Phi_i \ra $ written explicitly.

\begin{figure} 
 \epsfxsize=8cm
 \epsfbox{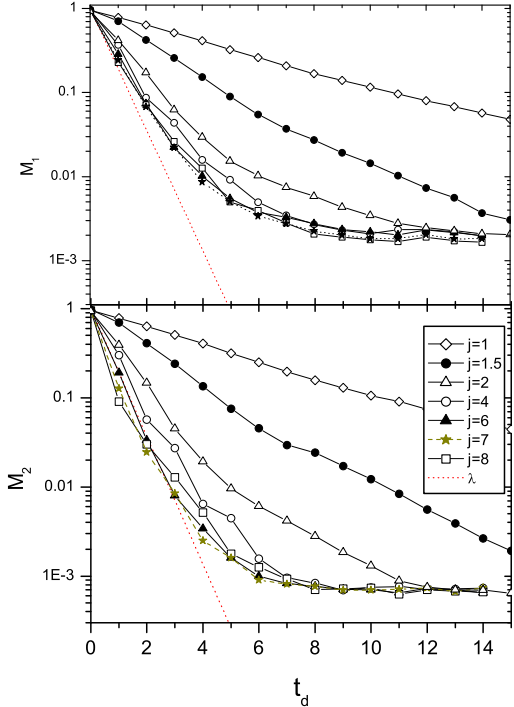}
 \vspace{-0.4cm}
\narrowtext \caption{ \label{fig-M1M2}
 Decay of averaged $M(t)$ of initial coherent states,
 from the golden rule regime ($j=1$) to the perturbation-independent regime, for $K=13.1$.
 The difference between $M_1$ and $M_2$ lies in the averaging procedure
 (see Eqs.~(23) and (24)).
 The dotted straight lines represent the exponential decay with the Lyapunov
 exponent $\lambda =1.65$.}
\end{figure}

 Numerical results for $M_1(t_d)$ (upper) and $M_2(t_d)$ (bottom)
 are shown in Fig.~\ref{fig-M1M2},
 with increasing $\kappa $ ($j$ from 1 to 8), when $K=13.1$.
 In calculating the two quantities, $t_s$ for each $M(t)$ of $j=1$
 and 1.5 was taken to be the first
 $t$ at which $M(t+1) <0.9$.
 For $j \ge 2$, $t_s$ was determined by the first
 $t$ at which $M(t) / M(t+1) > 1.5$.
 Only states satisfying $M(t_s) > 0.85$ were used in averaging.
 The reason of taking two methods in calculating $t_s$  is
 that $M(t)$ decays slowly at $j=1$ and 1.5.
 The number $N$ of coherent states taken for averaging is 1000.
 Results in Fig.~\ref{fig-M1M2} for $M_1(t_d)$ are in consistence with those in Ref.~\cite{JSB01},
 namely, the decay of $M_1(t_d)$ saturates at
 an exponential decay which is slower than the one with the Lyapunov exponent.
 However, when the large fluctuations in individual $M(t)$ are suppressed
 by the second averaging procedure,
 Fig.~\ref{fig-M1M2} shows that the decay of the averaged fidelity $M_2(t_d)$ saturates at
 an exponential decay that is close to the  Lyapunov decay.
 The deviation of the saturation decay of $M_1(t_d)$ from the one
 predicted by the semiclassical theory is caused by the large fluctuations
 in individual $M(t)$.
 In fact, this can also be seen roughly in Fig.~\ref{fig-singleMt}.

 \begin{figure} 
 \epsfxsize=8cm
 \epsfbox{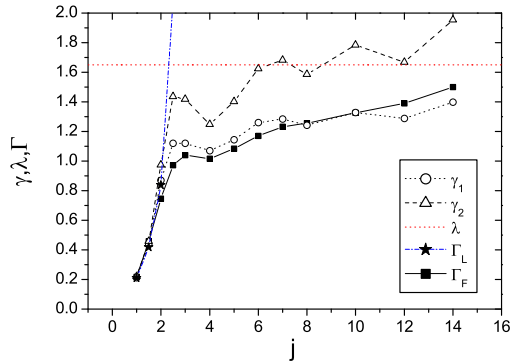}
 \vspace{-0.4cm}
 \narrowtext \caption{ \label{fig-gamma-sigma}
  Values of $\gamma_1$ and $\gamma_2$ of the exponential decay of $M_1(t_d)$
  and $M_2(t_d)$, respectively,  at different values of the perturbation parameter,
  together with $\Gamma_L$, the half width the averaged LDOS,
  and $\Gamma_F$, the averaged half-width of the best fitting Lorentzian form
  to the central part of $\rho_F(E) = |F_M(E)|$.
  The value of the Lyapunov exponent of the underlying classical dynamics,
  $\lambda =1.65$, is indicated by the horizontal dotted line.
  }
 \end{figure}

 To show the difference between the decay of $M_1(t_d)$ and that of $M_2(t_d)$ in
 Fig.~\ref{fig-M1M2}
 in a quantitative way,
 we calculate their decaying exponents $\gamma_1$ and $\gamma_2$,
 \be M_1(t_d) \propto {\rm exp}(-\gamma_1 t_d)
 \hspace{0.5cm}  M_2(t_d) \propto {\rm exp}(-\gamma_2 t_d), \label{gamma12} \ee
 the values of which are presented in Fig.~\ref{fig-gamma-sigma}.
 Numerically, $\gamma_i$ ($i=1,2$) are calculated by the best linear fitting
 to $\ln M_i(t_d)$, with the fitting lines fixed to the values of $\ln M_i(0)$ at $t_d=0$.
 The points  $t_d$ used in fitting are those
 satisfying $1 \ge  M_2(t_d) \ge  0.008$.
 In fig.~\ref{fig-gamma-sigma}, we see that, in the golden rule regime ($ j =1 $ and 1.5),
 $\gamma_1$ is close to $\gamma_2$ because of the small fluctuation in individual $M(t)$,
 and both of them are close to the half width of LDOS, $\Gamma_L$,
 as predicted by the random matrix theory.
 The difference of the three quantities becomes obvious, when $j \ge 2$.
 In the parameter regime $j \ge 6$, $\gamma_2$  fluctuates around
 the Lyapunov exponent $\lambda =16.5$,
 while $\gamma_1$ is obviously smaller than $\lambda $.
 The value of $\gamma_2$ at $j=14$ is obviously larger than
 the Lyapunov exponent.
 This may be due to the reason that the perturbation cannot be regarded as small
 at this perturbation value.
 In fact, when the perturbation is so strong that the classical perturbation
 theory breaks down, $M(t)$ may decay faster than the Lyapunov decay \cite{CLMPV02}.
 As discussed above,
 when $S$ is increased and we are more and more close to the classical limit,
 the fluctuations in $M(t)$ shown in Fig.~\ref{fig-singleMt}
 should become small and small,
 which would make $\gamma_1$ approach to $\gamma_2$.
 Indeed, we calculate the case
 of $S=1000$ ($\kappa $ reduced to half to those of $S=500$), where
 $\gamma_2$ are found to be fluctuating around $\lambda $ as well
 and $\gamma_1$  a little larger than the values at $S=500$.
 However, $\gamma_1$ of $S=1000$ are still obviously smaller than $\lambda $,
 showing that $S=1000$ is still not in the deep semiclassical regime.

 \subsection{Dependence of fidelity on initial states}
 \label{ISD}

\begin{figure}   
 \epsfxsize=8cm
 \epsfbox{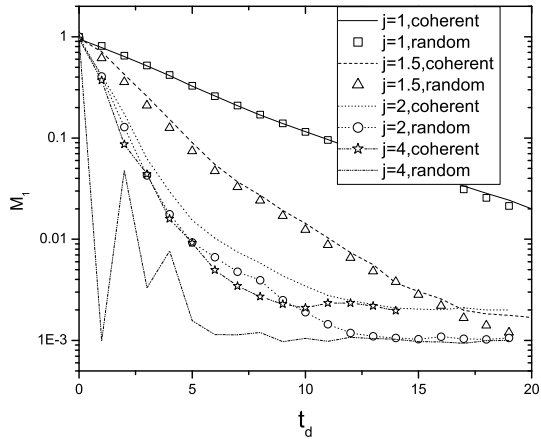}
 \vspace{-0.4cm}
 \narrowtext  \caption{  \label{fig-random-regular}
 Comparison of the decay of the averaged fidelity of initial coherent states
 and that of initial random states in Eq.~(\ref{randomIS}).}
 \end{figure}

 In different regimes of the perturbation parameter,
 the initial state can play different roles
 in influencing the behavior of the fidelity.
 In the Golden rule regime, the LDOS is narrow
 and, as discussed in Sect.~\ref{F_M},
 the random matrix theory can be used to predict the behavior of the fidelity.
 In this regime, there should be no difference between
 the decaying behavior of the fidelity of an initial coherent state
 and that of an arbitrary initial state,
 that is, the decay of the fidelity should be initial-state-independent.
 Indeed, in a study of the influence of sub-Planck scale structures \cite{Zurek01}
 on the fidelity in Ref.~\cite{JAB02},
 the averaged fidelity of the initial states
 \be |\Phi_0(T) \ra = {\rm exp}(-iH_0 T) |\Phi_0^c \ra , \ee
 where $|\Phi_0^c \ra $ are coherent states,
 has been found numerically to be independent on the parameter $T$ and
 almost the same as that of initial coherent states.
 We compare the averaged fidelity of initial coherent states
 with that of initial random states of the form
 \be | \Phi_0^{\rm random} \ra = \sum_{\alpha} c_{\alpha } | \alpha\ra , \label{randomIS}  \ee
 where $c_{\alpha}$ are random complex numbers
 satisfying the normalization condition.
 It is found that the former ones have `shoulders',
 while the latter ones do not.
 With the `shoulders' of the former ones subtracted,
 the decay of the corresponding averaged fidelity is found to be close to each other
 in the golden rule regime.
 Some of the results are plotted in Fig.~\ref{fig-random-regular},
 where 1000 initial states were used in averaging for each case.

\begin{figure} 
 \epsfxsize=8cm
 \epsfbox{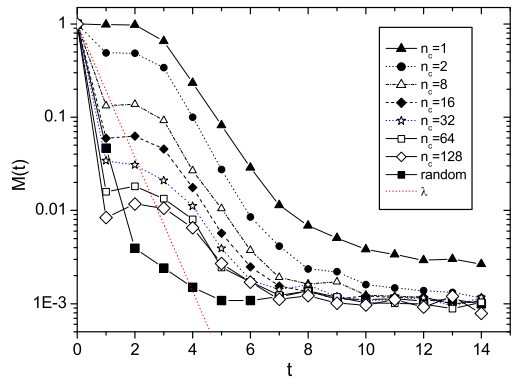}
 \vspace{-0.4cm}
 \narrowtext  \caption{
 The averaged fidelity with $|\Phi_0^{\rm rs}(n_c) \ra $, a
 random superposition of coherent states, being the initial states.
 The average is taken in the same way as for $M_1(t_d)$.
 The solid squares connected by the solid line
 show the averaged fidelity of initial random states
 in Eq.~(\ref{randomIS}).
 The dotted straight line indicates the exponential decay with the Lyapunov exponent $\lambda $.
 $j=6$ for the perturbation parameter.
 \label{fig-random} }
 \end{figure}

 In the perturbation-independent regime of the fidelity of initial coherent states,
 $M(t)$ of $|\Phi_0(T) \ra $ is found to
 be decaying  in part of the evolution time in a way similar to that of
 initial coherent states \cite{JAB02}.
 This implies that, although the system $H_0$ is classically chaotic,
 the time evolution operator ${\rm exp}(-iH_0 T)$ does not destroy the coherence
 possessed by a coherent state.
 Indeed, we find that random superpositions of coherent states defined by
 \be |\Phi_0^{\rm rs}(n_c) \ra = \sum_{n=1}^{n_c} e^{i\varphi_{n}} |z_n \ra , \ee
 where $\varphi_{n}$ are random phases and $|z_n\ra$ are coherent states
 in Eq.~(\ref{CS}),
 still have some coherence properties of the coherent states.
 The averaged fidelity with $|\Phi_0^{\rm rs}(n_c) \ra $ being initial states
 at the perturbation parameter $j=6$
 are presented in Fig.~\ref{fig-random}.
 We see that the averaged $M(t)$  with $n_c > 1$ has a
 rapid decrease between $t=0$ and 1,
 which is faster than the Lyapunov decay  for $n_c \ge 8$.
 This indicates that part of the coherence in the coherent components $|z \ra $
 of the initial states $|\Phi_0^{\rm rs}(n_c) \ra $  is destroyed
 by the random phases in the initial states.
 In fact, when $\la \alpha | \Phi_0 \ra $,
 the components of initial states in the eigenstates $|\alpha \ra $,
 can be treated as random numbers,
 the random matrix theory analysis predicts that
 the decay of the averaged $M(t)$ is controlled by the averaged LDOS \cite{JAB02},
 the half width of which is larger than the Lyapunov exponent at $j=6$ (Fig.~\ref{fig-gamma-sigma}).

 After the initial transient rapid decrease,
 the remanent coherence in the initial states $|\Phi_0^{\rm rs}(n_c) \ra $
 plays a role similar to that of initial coherent states.
 Indeed, the averaged $M(t)$ changes slowly in some short time intervals after $t=1$,
 which are analogues of the `shoulders' of the fidelity of initial coherent states,
 then, decreases exponentially
 in a way similar to that of initial coherent states,
 until they become close to the saturation value,
 which is about $1/(2S)$.
 The larger the value of $n_c$ is, the smaller the remanent coherence will be.
 The fidelity of random superpositions of $|\alpha \ra $ in Eq.~(\ref{randomIS})
 decays faster than the Lyapunov decay shown by the dotted straight line in Fig.~\ref{fig-random},
 as predicted by the random matrix theory.

 \subsection{Properties of $F_M(E)$ and relation to properties of $M(t)$}
 \label{numericalF_M}

 As shown in Sect.~\ref{F_M}, the Fourier transform of $M(t)$, the function $F_M(E)$,
 is directly determined by the states $|\alpha \ra $,
 $|\beta \ra $ and $|\Phi_0 \ra $.
 In this subsection, we study properties of the function $F_M(E)$ in the kicked top model
 numerically,
 in particular, its absolute value part $\rho_F(E)$
 and its phase part $\theta_F(E)$.
 Unless addressed explicitly, initial states $|\Phi_0 \ra $ studied in this subsection
 are coherent states.
 It will be shown that properties of the central part of
 $F_M(E)$ are closely related to the decaying rate and the
 width of the `shoulder' of the averaged fidelity $M(t)$.

\begin{figure} 
 \epsfxsize=8cm
 \epsfbox{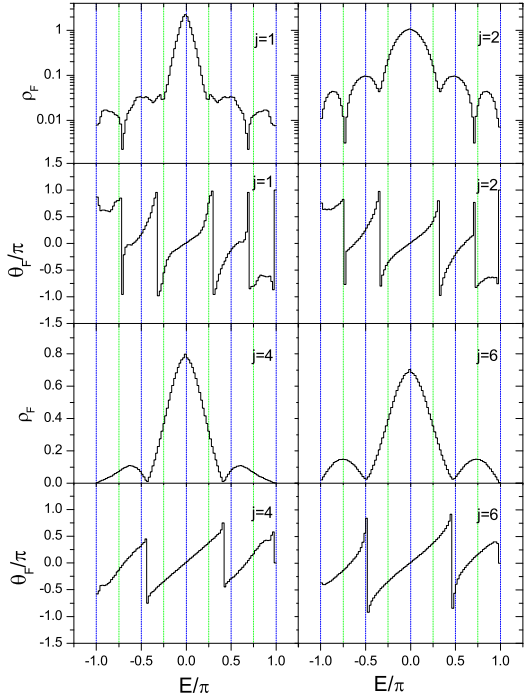}
 \vspace{-0.4cm}
\narrowtext \caption{ \label{fig-FM1246}
 Typical shapes of $\rho_F(E)$ and $\theta_F(E)$ of the same coherent state,
 for $j=1$, 2, 3, 4. $\rho_F(E)$ are plotted in the logarithm scale for $j=1$ and 2.  }
\end{figure}

 Typical shapes of $\rho_F (E)$ and $\theta_F(E)$
 in cases of different values of the perturbation parameter
 are shown in Fig.~\ref{fig-FM1246},
 where the same coherent state is used.
 For $\rho_F(E)$ of $j=1$ and 2, the values of $\rho_F$ are plotted in the logarithm scale.
 Each $\rho_F(E)$, symmetric with respect to $E=0$,
 is composed of three or more parts, with a main central part.
 A manifestation of the coherence in the coherent state $|\Phi_0 \ra $
 is that the corresponding phase functions $\theta_F(E)$
 have related divisions, as can be seen in Fig.~\ref{fig-FM1246}.
 (If $|\Phi_0 \ra $ is taken as a random state $|\Phi_0^{\rm random } \ra $
 in Eq.~(\ref{randomIS}), values of the function $\theta_F(E)$ will be
 close zero.)

\vspace{0.1cm}

\begin{figure} 
 \epsfxsize=8cm
 \epsfbox{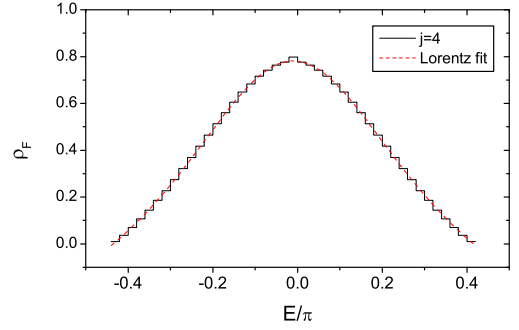}
 \vspace{-0.5cm}
\narrowtext \caption{ \label{fig-C-BWfit}
 Lorentzian fit to the central part of the $\rho_F(E)$ in the $j=4$ case in Fig.~\ref{fig-FM1246}. }
\end{figure}

 The central part of each $\rho_F$ is found to be fitted well
 by the Lorentzian form, with a width denoted by $\Gamma_F$,
 to be distinguished from $\Gamma_L$ for the width of the LDOS.
 See Fig.~\ref{fig-C-BWfit} for an example of the Lorentzian fit to
 the central part of the $\rho_F$
 in Fig.~\ref{fig-FM1246} of j=4.
 The Lorentzian form used in fitting is  $\rho_L(E)+a$,
 where $\rho_L(E)$ is the function on the right hand side of Eq.~(\ref{BW})
 and $a$ is a fitting constant.
 Therefore, the closeness of the central part of $\rho_F(E)$ to the fitting Lorentzian
 form does not predict directly an approximate exponential decay of $M(t)$ with an exponent $\Gamma_F$.
 In fact, the contribution of the central part of $F_M(E)$
  to $M(t)$ in Eq.~(\ref{Mt-FM}) gives both positive and negative results.
 The final positive results for $M(t)$ come out, only when the contributions of the
 other parts of $F_M$ are taken into account.
 However, the average values of $\Gamma_F$,
 which are shown by solid squares  in Fig.~\ref{fig-gamma-sigma},
 are found numerically to be close to
 the values of $\gamma_1$ for $M_1(t_d)$,
 in the whole parameter regime of $j$ from 1 to 14.
 When $S$ is taken equal to 1000, the change of the corresponding results of
 $\Gamma_F$ are found to be smaller than that of $\gamma_1$.
 Since $\gamma_1$ increases only a little, when $S$ is changed from 500 to 1000,
 it is not clear at the present stage
 whether or not $\Gamma_F$ can approach to $\gamma_1$ in the classical limit.

\begin{figure} 
 \epsfxsize=8cm
 \epsfbox{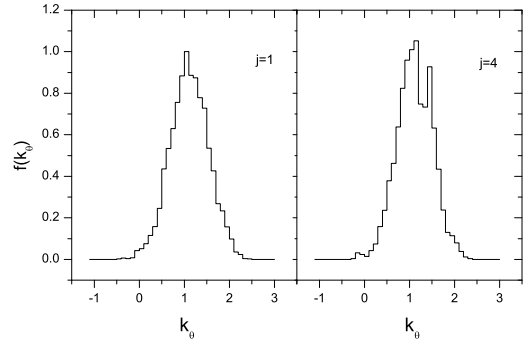}
 \vspace{-0.1cm}
\narrowtext \caption{ \label{fig-dis-kt}
 Distributions of $k_{\theta }$, the slope of the approximate linear
 behavior of $\theta_F(E)$ in the central region of $F_M(E)$. }
\end{figure}

\begin{figure} 
 \epsfxsize=8cm
 \epsfbox{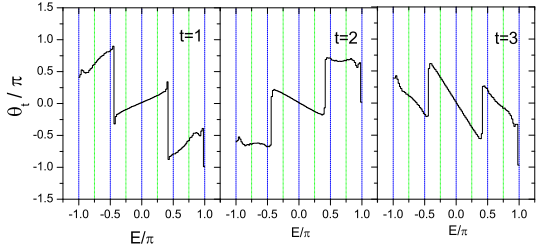}
 \vspace{-0.4cm}
\narrowtext \caption{ \label{fig-w-theta}
 Change of $\theta_t(E)$ with $t$, for the $\theta_F(E)$ of $j=4$
 in Fig.~\ref{fig-FM1246}. }
\end{figure}

 Numerical results for the phase function $\theta_F(E)$
 show that it is approximately linear
 in the region of $E$ corresponding to the central part of $\rho_F$
 (see Fig.~\ref{fig-FM1246} for examples),
 \be \theta_F(E) \approx k_{\theta } E, \label{ktheta} \ee
 where $k_{\theta }$ is the slope of the approximate linear behavior.
 Distributions of $k_{\theta }$
 are shown in Fig.~\ref{fig-dis-kt} for some perturbation parameters.
 The positions of the peaks of the distributions are close to $k_{\theta } =1$,
 while the distribution of $j=4$ has another small peak at $k_{\theta } =1.5$.
 To illustrate the influence of $\theta _F(E)$ on behaviors of $M(t)$ clearly,
 let us write $M(t)$ in the form
 \be M(t) = \int \rho_F(E) e^{i\theta_t(E)}  d E, \label{Mt-FM2} \ee
 where ${\theta_t(E)}= \theta_F(E) -Et$.
 Some examples of the behavior of $\theta_t(E)$ are shown in Fig.~\ref{fig-w-theta}.
 With increasing $t$, the approximate slope of $\theta_t(E)$ in the central region
 of $F_M(E)$  changes from positive to negative.
 For $t \ge  2$, the larger $t$ is, the steeper $\theta_t(E)$ will be.
 As shown below, this feature of $\theta_t(E)$ can be used
 to estimate the width of the `shoulder' of $M(t)$.

\begin{figure} 
 \epsfxsize=8cm
 \epsfbox{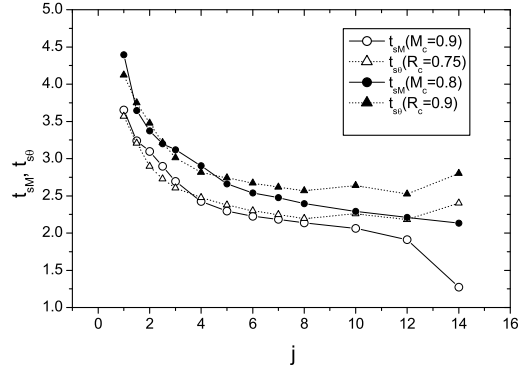}
 \vspace{-0.4cm}
\narrowtext \caption{ \label{fig-td}
 Two examples of the relationship between $t_{\rm sM}$ and $t_{s\theta }$.
 They are close to each other, when $j$ is not too large, namely, $1 \le j \le 8$. }
\end{figure}

\begin{figure} 
 \epsfxsize=8cm
 \epsfbox{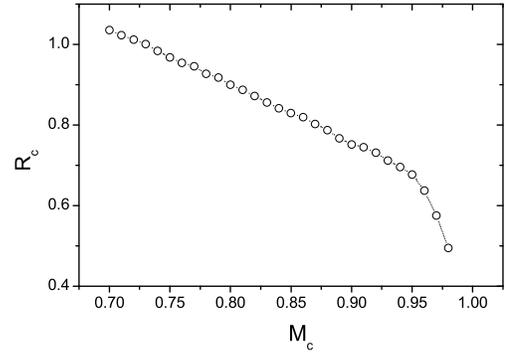}
 \vspace{-0.4cm}
\narrowtext \caption{ \label{fig-MRc}
 Dependence of $R_c$ on $M_c$, required by the smallest value of (\ref{sumtm}).}
\end{figure}

 In order to give an estimation of the `shoulder' width of $M(t)$,
 we note that the exponential decay of $M(t)$ should appear when $\theta_t(E)$ is steep enough,
 specifically, when $\theta_t(-W_F/2)$ is comparable to $\pi $,
 where $W_F$ is the width of the region of $E$
 occupied by the main central part of $\rho_F$.
 Quantitatively, we use $t_{{\rm s}\theta }$ to denote
 the average of the largest $t$ satisfying
 \be |k_{\theta }-t| \frac{W_F}2 \le R_c  \pi , \label{tstheta} \ee
 where $R_c$ is a quantity used to show the closeness to $\pi $.
 The average width of the `shoulders' of $M(t)$, denoted by $t_{\rm sM}$,
 are calculated from the averaged $M(t)$ directly.
 Since $t$ has integer values only in the fidelity $M(t)$ of the kicked top model,
 we use the linear interpolation in calculating $t_{\rm sM}$ by
 $M(t_{\rm sM})=M_c$, where $M_c$ is a quantity measuring
 the lower border of the `shoulder' of the averaged $M(t)$.
 In studying the relation between $t_{\rm sM}$ and $t_{{\rm s}\theta }$,
 for a given value of $M_c$,
 the value of $R_c$ is determined by the smallest value  of
 \be \sum_{j=1}^8 (t_{{\rm s}\theta } - t_{sM})^2, \label{sumtm} \ee
 in the parameter regime $1 \le j \le 8$.
 Figure \ref{fig-td} shows two examples of $M_c=0.9$ with $R_c \simeq 0.75$
 and $M_c=0.8$ with $R_c \simeq 0.9$,
 where we see that $t_{{\rm s}\theta } \approx t_{\rm sM}$ for $j $ between 1 and 8.
 The difference between $t_{{\rm s}\theta }$ and  $t_{\rm sM}$
 becomes increasingly large, when $j$ is larger than 8,
 which is related to the fact that the perturbation $\kappa V$ is not
 quite small for $j > 8$.
 Variation of $R_c$ with respect to $M_c$ is presented in Fig.~\ref{fig-MRc},
 where $R_c$ is seen to be almost linear with $M_c$ for $M_c$ between 0.7 and 0.95.

\section{Conclusions and Discussions}
\label{Conclusion}

 In this paper,
 we study the crossover of the fidelity
 from the golden rule regime to the perturbation-independent
 decay regime.
 We show that the deviation of the perturbation-independent decay of the
 averaged fidelity from the prediction of the semiclassical theory
 is due to large quantum fluctuations in individual fidelity
 caused by the coherence in the initial coherent states.
 When the quantum fluctuations are suppressed by an appropriate averaging procedure,
 the perturbation-independent decay of the fidelity is found
 to be close to the semiclassical prediction.

 We also show that when the effective Planck constant is not small enough
 to guarantee the validity of the semiclassical theory,
 the crossover does not occur at a point of the perturbation parameter,
 but takes place in a region of the parameter (Fig. 3).
 The crossover in this case can be analyzed by neither the semiclassical theory,
 because the effective Planck constant is not small enough,
 nor the random matrix theory,
 since the perturbation-independent decay of the fidelity is initial-state-dependent.

 Furthermore, a function $F_M(E)$ that is in fact the Fourier
 transform of the fidelity, is introduced to study the fidelity in
 the crossover region. This function is determined directly by the
 initial state and the eigenstates of the two Floquet operators
 concerned.
 Numerically, for initial coherent states,
 the absolute value part and the phase part of the function $F_M(E)$ are found
 to have related division into several sub-parts,
 which is a manifestation of the coherence possessed by the initial coherent states.
 In the whole crossover region,
 properties of the central part of $F_M(E)$ are
 found to be closely related to both the exponent of the
 exponential decay
 of the averaged fidelity and the length of the short initial time period, within which the
 fidelity almost does not change due to the coherence possessed by the initial
 coherent state and the smallness of the perturbation in the classical limit.

 The function $F_M(E)$ is studied mainly numerically in this paper.
 A detailed analytical analysis of the function,
 e.g., by making use of its expression as the correlation function of
 $f_m(\epsilon)$
 given in Sect.~\ref{F_M},
 would supply further understanding of the behaviors of the fidelity,
 in analogue to the relation between the LDOS and the survival probability.
 Information of the coherence in a coherent state may be hidden in its expanding
 coefficients in the eigenstates of classically chaotic  systems.
 Numerical results in this paper show that the function $F_M$ supplies a useful
 method of extracting such information.
 We would like to mention that some other quantities discussed in this paper
 may serve this purpose as well in other situations, e.g.,
 the quantity $A_{\alpha \beta }$ introduced in Sect.~\ref{F_M},
 which is irrelevant to the arbitrary phases that can be given to
 the initial coherent state and the quasi-energy eigenstates of the two
 classically chaotic systems.

\acknowledgments

 The authors are grateful to G.~Casati for valuable discussions.
 The work was supported in part by the Academic Research Fund of the
 National University of Singapore and the DSTA of Singapore. We
 also thank Dr. Wang Jiao for providing the programme for
 diagonalization of unitary matrix.

\end{multicols}

\end{document}